\documentclass[article]{iucr}              
\usepackage{graphicx} 
\usepackage{bm} 
\usepackage[intlimits]{amsmath}
\usepackage{esint}
\usepackage{caption}

\begin{document}                  



\title{Measurement and modeling of polarized specular neutron reflectivity in large magnetic fields}

\cauthor[a]{Brian B.}{Maranville}{brian.maranville@nist.gov}{}

\author[a]{Brian J.}{Kirby}
\author[a]{Alexander J.}{Grutter}
\author[a]{Paul A.}{Kienzle}
\author[a]{Charles F.}{Majkrzak}

\aff[a]{%
 NIST Center for Neutron Research  
 100 Bureau Drive, Gaithersburg MD USA 20899
}%

\author[b]{Yaohua}{Liu}
\aff[b]{%
 Quantum Condensed Matter Division, Oak Ridge National Laboratory, Oak Ridge TN USA 37831
}

\author[c]{Cindi L.}{Dennis}
\aff[c]{
 NIST Material Measurement Laboratory
 100 Bureau Drive, Gaithersburg MD USA 20899
}









\maketitle                        

\ifx \circled \undefined
  \def \circled[#1]{\raisebox{.5pt}{\textcircled{\raisebox{-.9pt} {#1}}}}
\fi

\begin{synopsis}
Procedure for polarized neutron reflectometry when the Zeeman corrections are 
significant, which occurs when both the magnetic anisotropy and applied magnetic
field are significant.  Calculations and recommended procedure in an example 
system are provided.
\end{synopsis}

\begin{abstract}
The presence of a large applied magnetic field removes the degeneracy 
of the vacuum energy states for spin-up and spin-down neutrons.  For polarized
neutron reflectometry, this must be included in the
reference potential energy of the Schr\"odinger equation that is used 
to calculate the expected scattering from a magnetic layered structure.
For samples with magnetization that is purely parallel or antiparallel to the
applied field which defines the quantization axis, there is no mixing of the spin
states (no spin-flip scattering) and so this additional potential is constant throughout
the scattering region.  When there is non-collinear magnetization in the sample 
however, there will be significant scattering from one spin state into the other
and the reference potentials will differ between the incoming and outgoing wavefunctions,
changing the angle and intensities of the scattering.  The theory of the scattering and
recommended experimental practices for this type of measurement are presented, 
as well as an example measurement.
\end{abstract}

\section{Introduction} \label{intro}

Polarized specular neutron reflectometry measurements require at least a small
magnetic field to be applied throughout the measurement apparatus, in order to maintain
a well-defined neutron quantization axis.  In addition, a larger field is 
often applied at the sample position in order to manipulate the magnetic state of
the sample \cite{PNRMajkrzakChapter}.  The difference 
in the Zeeman energy for a spin-up vs. a spin-down neutron can lead to observable shifts in
both the angle and intensity of scattering 
for even modest applied fields (10s of mT) when spin-flip scattering is 
appreciable; when the spin-flip cross-section is small compared to the 
non-spin-flip, the corrections remain small. 

This so-called Zeeman shift in the spin-flipped reflected neutrons 
was first described by Felcher et al. \cite{felcher1995zeeman}, and 
observed by many others \cite{felcher1996observation}; in  Ref. \cite{kozhevnikov2012data} 
a clear description of the geometry of the incident and scattered beams is presented.
The reflectivity calculation formalism including the Zeeman term is briefly described
in \cite{vandeKruijs2000189, PhysRevB.83.174418}, 
but to our knowledge a detailed description of the calculation is not available in the 
literature, nor has such a calculation been incorporated into commonly-used modeling software.


These shifts are not a major concern in many experiments \cite{PhysRevB.83.174418} 
because the effect is significant only when there is both a large applied field and 
strong spin-flip scattering.  At low fields the corrections are minimal, and at 
high fields the magnetization tends to align parallel to the applied field, 
so there is insignificant spin-flip scattering.  
However, there are important cases where accounting for the Zeeman shift is 
necessary for appropriately measuring and analyzing data.
A technologically relevant example is the study of high anisotropy magnetic material 
used in advanced data storage applications \cite{PhysRevB.83.174418}. 
In such cases the sample magnetization can be non-collinear with even large applied fields.


In this paper we will address the requirements for setting up a measurement in a large field
in the case where the spin-flip scattering is not negligible; we present the changes that
need to be made to a commonly-used existing computer algorithm (implemented in gepore.f \cite{PNRMajkrzakChapter})
in order to correctly calculate the scattering, and we present recommended practices for
performing the measurements when the applied magnetic field $\vec H$ and magnetization $\vec M$ 
are both large, and not parallel to each other.
This implies a large magnetic anisotropy in the system.  We
take advantage of the large shape anisotropy in a thin film of a soft magnetic 
material in the example experiment section of this paper to clearly show the effects we are discussing.

We must also address the meaning of the word ``specular''; in many texts on reflectivity
the definition is given that the angle of incidence equals the angle of reflection, or 
that the out-of-plane component of the momentum of the incoming beam $k_{z,\mathrm{in}}$
is equal in magnitude to that of the outgoing reflected beam $k_{z,\mathrm{out}}$.
Here we will use a more functional definition based on the
momentum transfer $\vec Q \equiv \vec k_\mathrm{in} - \vec k_\mathrm{out}$; we define the reflectivity
as specular on the condition that the in-plane momentum transfers $Q_x = 0$ and $Q_y = 0$,  
so that the momentum transfer $\vec Q \equiv Q_z \hat z$ (perpendicular to the surface)
as is expected when reflecting from planar layered samples.

As we will demonstrate, some of the kinetic energy along $\hat z$
is traded for potential energy during a spin-flip process, so the earlier definitions 
do not apply in this circumstance, while $\vec Q$ remains strictly out-of-plane.

\section{Boundary conditions}

Starting with the general Schr\"odinger equation for a neutron with spin $\frac{1}{2}$:
\begin{equation}
\label{eq:general_schrodinger}
\left[
  - \frac{\hbar^2}{2m}
    \nabla^2 
    \hat 1
  + \hat V(\mathbf{r})
  - E  \hat 1
\right]
  \begin{pmatrix}
    \psi^+(\mathbf{r})  \\
    \psi^{-}(\mathbf{r}) 
  \end{pmatrix}
= 0
\end{equation}
where $\psi^{\pm}$ is the spin-dependent wavefunction for the neutron, 
$\hat 1 = \begin{pmatrix} 1 & 0 \\ 0 & 1 \end{pmatrix}$,
$\nabla^2$ is the Laplacian (spatial second derivative) and
the hatted components indicate a Pauli spin matrix with $z'$ as
the quantization axis.  We use the notation $z'$ for coordinates in the 
spin quantization reference frame to distinguish it from the scattering geometrical
reference frame where $z$ is defined to be the surface normal direction for the
planar sample, and there is no requirement that $\hat z \parallel \hat z'$.
The potential of the particle is made up of a 
scalar nuclear potential $V_\mathrm{nuc}$ and a magnetic potential due to 
the field $\mathrm{B}$: 
\begin{equation}
	\hat V = \mu_N \boldsymbol{\sigma} \cdot \mathbf{B} + V_\mathrm{nuc} \hat 1
\end{equation}
where
\begin{equation}
  	 \boldsymbol{\sigma} \cdot \mathbf{B} =
  	 	  \begin{pmatrix} 0 & 1 \\ 1 & 0 \end{pmatrix} B_{x'} +
  		  \begin{pmatrix} 0 & -i \\ i & \phantom{-}0 \end{pmatrix} B_{y'} +
  		  \begin{pmatrix} 1 & \phantom{-}0 \\ 0 & -1 \end{pmatrix} B_{z'} 
\end{equation}

In the ``prepared'' spin-polarized beam, we define the direction of the guide field to be 
$\hat z'$, so there are no off-diagonal elements to the potential above 
(because $B_{x'} \equiv B_{y'} \equiv 0$) and the equation decouples into 
two linear equations for potentials with $V=V_\mathrm{nuc} \pm \mu_N B_{z'}$.


When the beam enters the fronting medium with non-negligible $B$, there is no 
physical restriction on the direction of $\vec B$, but from an experimental
design perspective we note that if the magnetic field in the fronting  
medium is not parallel to the applied laboratory field direction $(\hat z')$, 
i.e. there is a non-zero $B_{x'}$ or $B_{y'}$ component to the
field in this region, the wavefunction will be angularly split due to the field-dependent
difference between $k_{F,x}^+$ and $k_{F,x}^-$.
The mutual coherence of the
two resulting beams will be impractical to calculate over the macroscopic distances
the beam will then travel after being split.

This is not to be confused with the angular splitting which occurs as the
the beam interacts with the horizontal layers of the sample, which is what 
is usually being discussed when describing reflectivity, and which 
is fully taken into account in the calculations below.  

\begin{figure}
	\caption{neutron entering field and nuclear potential region from the side.}
	\includegraphics[width=1.0\linewidth]{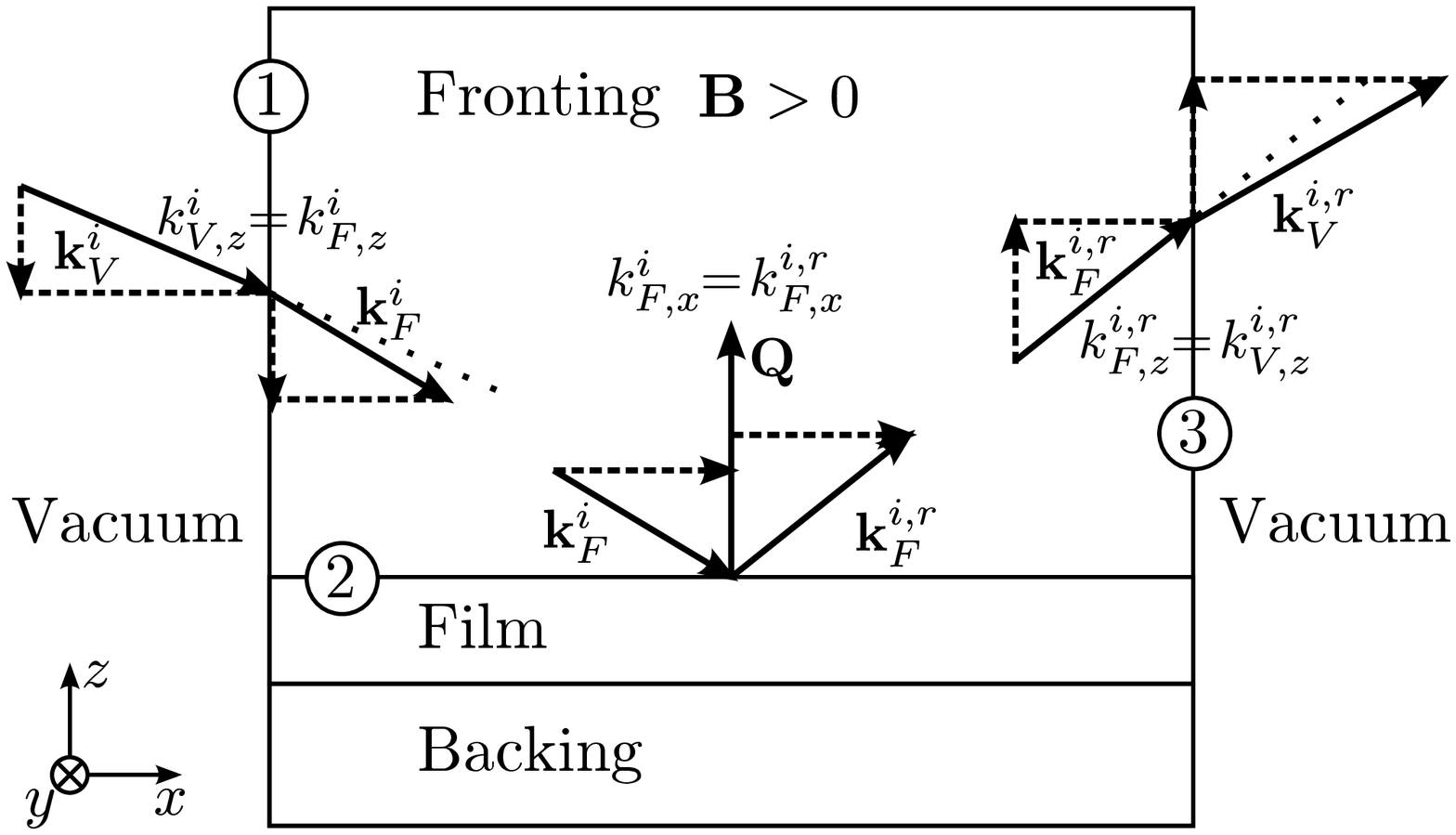}
	\label{fig:boundaries}
\end{figure}

Now restricting ourselves to the case in which the $B$-field in the fronting
medium is parallel to the guide field outside the fronting medium, we can fully
describe the interaction of the neutron with the sample as in Fig. \ref{fig:boundaries}.

The incident neutron is ``prepared'' in either the $I^+$ or $I^-$ spin state with techniques 
described elsewhere \cite{dura2006and, PNRMajkrzakChapter} and 
neglecting the contribution of a very small magnetic guide field, 
the total energy of both states is nearly the same for the same $\mathbf{k}_V$,
and is equal to the kinetic energy:
\begin{equation}
	E^+_{V,xyz} \approx E^-_{V,xyz} \approx \frac{\hbar^2}{2m}\left(k_{V,x}^2 + k_{V,y}^2 + k_{V,z}^2\right)
\end{equation}
We note that the problem as defined has no $y$-dependence;  there are
no interfaces along that direction (out of the page of the figure) 
and so the solution for the wave equation along $y$ is for a plane
wave $e^{i k_y y}$ with constant kinetic energy that can be included in the total 
energy $E$ and the problem treated as a 2-dimensional Schrodinger equation
in $x$ and $z$, with
\begin{equation}
  \begin{array}{l l}
    E_{V,xz}^{\pm} &\approx \frac{\hbar^2}{2m} \left(k_{V,x}^2 + k_{V,z}^2\right)  \\[0.3em]
	{}&=E_{V,xyz}^{\pm} - \frac{\hbar^2}{2m} k_{V,y}^2
  \end{array}
\end{equation}

When the neutron enters the fronting medium at boundary \circled[1] 
as shown in Fig. \ref{fig:boundaries},
the potential energy changes in a spin-dependent way,  so that
\begin{equation}
	\label{eq:E+}
	E^+_{F,xz} = \frac{\hbar^2}{2m}\left[
	(k_{F,x}^{+})^2 +(k_{F,z}^{+})^2 + 4\pi(\rho_{F,N} + \rho_{F,B})
	\right]
\end{equation}
\begin{equation}
	\label{eq:E-}
	E^-_{F,xz} = \frac{\hbar^2}{2m}\left[
		(k_{F,x}^{-})^2 +(k_{F,z}^{-})^2 + 4\pi(\rho_{F,N} - \rho_{F,B})
	\right]
\end{equation}
where the notation $\mathbf{k}_{F}^{i}$ indicates the wavevector in the 
medium ($F$ for fronting and $V$ for vacuum) with spin state $i$ ($+$ or $-$).
The nuclear scattering length density of the fronting medium
$\rho_{F,N}$ depends on the isotopic composition of the medium,
while $\rho_{F,B}$ is the magnetic scattering length density, which can be calculated from
the magnetic field in that layer by
\begin{equation}
    \label{eq:Btorho}
	\rho_B = \frac{2 \mu_n m_n|\vec B|}{4 \pi \hbar^2} \approx B \times 2.31605 \times 10^{-6} \textrm{\AA}^{-2} \textrm{T}^{-1} 
\end{equation}
where $\mu_n, m_n$ are the magnetic moment and mass of the neutron, respectively
and $B$ is the magnetic field in the fronting medium (in Teslas). 

Since the magnetic field inside the vertical boundary is parallel to (though possibly much
bigger than) the field outside, the $(+)$ or $(-)$ spin state inside the vertical boundary 
matches the prepared state.
Also, by symmetry $k_z$ must be preserved across the vertical boundary
\circled[1], between 
the vacuum and the fronting medium, so $k_{V,z} = k_{F,z}$.  Since
$E_{V,xz}^+ = E_{F,xz}^+$ as well, this means that 
\begin{eqnarray}
	(k_{F,x}^{+})^2 = (k_{V,x})^2 - 4\pi(\rho_{F,N} + \rho_{F,B}) \\ \nonumber
	(k_{F,x}^{-})^2 = (k_{V,x})^2 - 4\pi(\rho_{F,N} - \rho_{F,B})
\end{eqnarray}
which changes the angle of the neutron beam inside the fronting
medium (this is refraction, as indicated by the shortened $k_x$ on the right
side of boundary \circled[1] in Fig. \ref{fig:boundaries}).
The energy trade in $k_x$ is reversed when
the neutron exits the fronting at boundary \circled[3]; the $k_{V,x}$ on 
the right is the same as it is on the left.  
This is not in general true
for $k_{V,z}$, as we will see.

Now we consider the next set of boundaries in the problem: 
the horizontal interfaces of the sample under investigation; the
first of these is the top interface \circled[2] between the fronting
and the sample.
When the neutron interacts with this structure, it is possible to have a spin-flip event,
so we introduce a second indicator $r$ (for \emph{reflected}) 
in the notation $\mathbf{k}_{F}^{i,r}$ for
the spin state of the outgoing neutron (still in the fronting medium).  
We retain the indicator $i$ for the 
\emph{incident} neutron spin state because this determines the energy in the fronting,
as described above.

We are considering the standard specular reflectometry case where the sample
under investigation is homogenous in-plane; so  
while $k_z$ was conserved across boundary \circled[1], now $k_x$ is conserved
across the boundaries like \circled[2], so
\begin{eqnarray}
	\label{eq:kx_conserved}
	k_{F,x}^{+,-} = k_{F,x}^{+,+}  = k_{F,x}^{+} \\ \nonumber
	k_{F,x}^{-,+} = k_{F,x}^{-,-} = k_{F,x}^{-}
\end{eqnarray} 
and because the total energy of the neutron is conserved during elastic scattering, we can write
\begin{equation}
  \begin{array}{l l}
	E_{F,xz}^{+,-} &= \frac{\hbar^2}{2m} \left[
	 (k_{F,x}^{+,-})^2 +(k_{F,z}^{+,-})^2 + 4\pi(\rho_{F,N} - \rho_{F,B})
	\right] \\ [0.3em]
	{} &= E_{F,xz}^{+}
  \end{array}
\end{equation}
Subtracting this from Eq. \ref{eq:E+} gives
\begin{equation}
	\label{eq:pm_shift}
	(k_{F,z}^{+,-})^2 = (k_{F,z}^{+})^2 + 8\pi\rho_{F,B}
\end{equation}
In a similar fashion for the $(-)$ state, we can obtain
\begin{equation}
	\label{eq:mp_shift}
	(k_{F,z}^{-,+})^2 = (k_{F,z}^{-})^2 - 8\pi\rho_{F,B}
\end{equation}
While the non-spin-flipped neutrons are not shifted:
\begin{eqnarray}
	(k_{F,z}^{+,+})^2 = (k_{F,z}^{+})^2 \\ \nonumber
	(k_{F,z}^{-,-})^2 = (k_{F,z}^{-})^2
\end{eqnarray}

At the next boundary \circled[3] where the neutrons exit the fronting material
and go back into the laboratory environment (vacuum) $k_z$ is again 
conserved by symmetry, as it was at \circled[1], so the shift in the spin-flipped
$k_z$ is carried across this boundary ($k_{V,z}^{i,r} = k_{F,z}^{i,r}$ for all $i,r$.)

The difference between $k_{V,z}^{+,-}$ and $k_{V,z}^{+,+}$ leads to a different 
propagation direction for the spin-flipped neutron; this measurable angular shift
is referred to as the Zeeman splitting.

There are values of $k_{F,z}^-$ for which $(k_{F,z}^-)^2 < 8\pi\rho_{F,B}$ and 
therefore the calculated momentum squared for the spin-flipped reflection $(k_{F,z}^{-,+})^2$ is
negative, so that $k_{F,z}^{-,+}$ becomes purely imaginary.  
The calculated amplitude for this reflection is valid at the interface, but this is
an evanescent wave that decays as it moves away from the sample.  The value of the 
measured reflectivity corresponds to the amplitude at the detector, and thus is effectively zero
in this case.

\subsection{Details of magnetic field geometry}
In the above discussion, the transition from vacuum with zero applied field
to a high-field region (also with a possibly non-zero nuclear scattering length density)
was described as a sharp boundary perpendicular to the sample plane (along $x$).
In that case, the momentum along $z$ is unchanged by the transition: 
$k_{F,z} = k_{V,z}$, and energy conservation leads only to a change in $k_x$:
$(k_{F,x}^+)^2 + 4\pi(\rho_{F,N} + \rho_{F,B}) = k_{V,x}^2$.

In real laboratory environments the magnetic field transition is not
as abrupt as what is shown in Fig. \ref{fig:boundaries}, and the direction is 
not perfectly defined, though typically the applied magnetic field is 
realized in a small volume centered on the sample and the field
gradient experienced by the probe neutron is to first order radial with respect
to the sample.
Since for any gradient potential the 
momentum components perpendicular to the gradient direction are conserved throughout
the interaction with the potential, the
abruptness of the transition is irrelevant and only the direction is important.

So, compared to a more realistic radial magnetic potential gradient 
parallel to the neutron momentum we
expect that by using our simplified rectangular boundary conditions (where the sharp
gradient at that boundary is along $\hat x$ and is nearly but not quite parallel to $\vec{k}_\mathrm{(in)}$) 
we introduce an error in the calculated $(k_{F,z}^\pm)^2$ proportional to  
$\sin^2 \delta$, where $\delta$ is the angle between the normal to the rectangular boundary
and $\vec{k}_\mathrm{(in)}$.  Because of the right angle between that boundary and the film surface, 
it ends up that 
$\delta$ coincides with the incident angle $\theta_\mathrm{in}$ of the neutron on the film surface.

At the small angles ($\theta_\mathrm{in}<$  6 deg.) commonly seen for the
incident angle during a reflectometry measurement, this results in a maximum correction to
$(k_{F,z}^\pm)^2$ from the model proposed above, on the order of 1 percent of $\pm 4\pi\rho_B$ 
(with the opposite correction made to $E_{F,z}^\pm$.)
At the even smaller angles ($\theta_\mathrm{in} \approx$ 0.5 deg.) near the critical edge where these
shifts might affect the modeling, the correction is just 0.01 percent of the magnetic
scattering length density.
For this reason in many cases it is a reasonable approximation that all the of the kinetic
energy shift in the Fronting region prior to the sample is along the $x$-direction, as defined by
the sample coordinate system in Fig. \ref{fig:boundaries}.

\section{Calculation of the spin-dependent reflectivity}
\subsection{1d Schr\"odinger equation}
Again considering the region between \circled[1] and \circled[3]
as above, we can calculate the reflectivity of the horizontally-layered 
structure there by reducing the Schr\"odinger equation to a 
single spatial dimension $z$ and solving with the boundary conditions
laid out above.
Since the potential is constant as 
a function of $x$ in this region (as it is for $y$ everywhere,) 
and $V(\mathbf{r}) = V(z)$, the one-dimensional plus spin 
Schr\"odinger equation for the neutron is then \cite{PNRMajkrzakChapter},
\begin{equation}
\label{eq:1d_schrodinger}
\left[
  - \frac{\hbar^2}{2m}
    \frac{\partial^2}{\partial z^2} 
    \hat 1
  + \hat V(z)
  - E_{F,z}^i  \hat 1
\right]
  \begin{pmatrix}
    \psi^{i,+}(z)  \\
    \psi^{i,-}(z) 
  \end{pmatrix}
= 0
\end{equation}
where 
\begin{equation}
  \begin{array}{l l}
	\hat V(z)\!\! &=
	\begin{pmatrix}
        V^{++}(z) & V^{+-}(z) \\ 
        V^{-+}(z) & V^{- -}(z)
    \end{pmatrix} \\ [1.2em]
    {} &= \displaystyle \frac{4\pi\hbar^2}{2m}\!\! 
    \begin{pmatrix}
        \rho_N + \rho_{Bz'} & \rho_{Bx'} - i\rho_{By'} \\ 
        \rho_{Bx'} + i\rho_{By'} & \rho_N - \rho_{Bz'}
    \end{pmatrix}(z)
 \end{array}
\end{equation}
and we fold the constant kinetic energy along $x$ into $E$ 
as we did for $y$ before:
\begin{equation}
	E^i_{F,z} = E^i_{F,xz} - \frac{\hbar^2}{2m} k_{F,x}^2 
\end{equation}

$E_{F,z}^i$ depends on the spin state of the incident neutron
as well as the potential in the fronting medium, as
\begin{equation}
	\label{eq:fronting_energy}
	E_{F,z}^\pm = \frac{\hbar^2}{2m} \left[4\pi(\rho_{F,N} \pm \rho_{F,B}) + (k_{V,z})^2\right]
\end{equation}

A set of solutions to Eq. \ref{eq:1d_schrodinger} is laid out in \cite{PNRMajkrzakChapter},
as (except now keeping track of the polarization $i$ of the incident state)
\begin{eqnarray}
\label{eq:psi_z_definition}
	\psi^{i,+}(z) = \sum_{j=1}^4 C^i_j e^{S^i_j z} \\ \nonumber
	\psi^{i,-}(z) = \sum_{j=1}^4 \mu_j C^i_j e^{S^i_j z}
\end{eqnarray}
where
\begin{eqnarray}
\label{eq:propagation_constants}
	S_1^i &=& \sqrt{4\pi(\rho_N + \rho_B) - \frac{2m}{\hbar^2}E_{F,z}^i  } \\ \nonumber
	S_2^i &=& -S_1^i	\\ \nonumber
	S_3^i &=& \sqrt{4\pi(\rho_N - \rho_B) -  \frac{2m}{\hbar^2}E_{F,z}^i } \\ \nonumber
	S_4^i &=& -S_3^i
\end{eqnarray}
\begin{equation}
\label{eq:mu_definition}
\begin{array}{l}
  \displaystyle \mu_1 = \mu_2 = \frac{ B + B_{x'}+iB_{y'} - B_{z'}}
    { B+B_{x'}-iB_{y'} + B_{z'}} \\[1.2em]
  \displaystyle \mu_3 = \mu_4 = \frac{ -B + B_{x'}+iB_{y'} - B_{z'}}
    { -B+B_{x'}-iB_{y'} + B_{z'}} 
\end{array}
\end{equation}
and the $C_j^i$ are the complex coefficients of the 4 components.

Within the fronting medium $F$ the propagation constants $S$ are equal to simply the
incident wave value $ik_{F,z}$, since the potentials $\rho$ cancel between 
Eqs. \ref{eq:fronting_energy} and \ref{eq:propagation_constants} for the 
incident beams $I^+, I^-$.

When the external magnetic potential is negligible, the $E$ in the above equations is
the same for both incident beam polarizations, but 
in general, $E^+_{F,z} \neq E^-_{F,z}$ for a sufficiently large field.
Because of this, if we measure reflectivity at the same $k_{V,z}$ for 
both the $I^+$ and $I^-$ states, we have to distinguish between polarization states
for the incoming beam.

This distinction based on the Zeeman energy of the neutron in the fronting medium is 
the basis for a small but critical change to the existing computer 
codes for calculating reflectivity (see gepore.f in \cite{PNRMajkrzakChapter}), where 
the term proportional to $E$ is set to $Q^2/4 - 4\pi\rho_{F,N}$ 
(for $Q\equiv 2k_{V,z}$), which accounts for only the kinetic and nuclear potential energy
in the fronting medium;  this gives the correct answer for any case except when
the Zeeman term is appreciable, so
we will use 
$\frac{2m}{\hbar^2}E_{F,z}^\pm$ instead, which includes the kinetic, nuclear and magnetic
energies in the fronting medium appropriate for the relevant incident spin state.  



Also in the previous code, Eq. \ref{eq:mu_definition}
for the ratio of $\psi^-$ to $\psi^+$ components is substituted with 
\begin{eqnarray}
	\mu_1 &=& \mu_2 = e^{i \theta_{\vec M}} \\ \nonumber
	\mu_3 &=& \mu_4 = -e^{i \theta_{\vec M}}
\end{eqnarray}
where $\theta_M$ is the in-plane $(x,y)$ angle, 
with the underlying, implicit assumptions that the contribution to $\vec B$ from
$\vec H_\mathrm{applied}$ is negligible and that the net $B_z$ (out of the sample plane) is zero.
These assumptions are quite reasonable for low values of $H$ 
even when there is a large perpendicular
magnetization, because for thin-film samples the demagnetization field
$|\vec H_D| = H_{Dz} \approx -M_z$ almost completely cancels the contribution of
the net perpendicular component $M_z$ to $B_z$ 
(because $\vec B = \mu_0[ M + \vec H_\textrm{applied} + \vec H_D + \ldots]$)
\footnote{Of course, the demagnetizing field does not exactly cancel the magnetic field
along $z$, and there is a non-zero
$\vec B$ field at large distances from the sample
(measurable with a magnetometer) 
which is proportional to volume integral of $\vec M$.
In the thin-film geometry, the surface to volume ratio goes to infinity,
and this is why there is effectively zero $B_\perp$ at the surface }

Now that we are including the effects of an arbitrary external field however,
we must include $B_{z} \approx H_{z}$ and return the more general 
equation (\ref{eq:mu_definition}) for $\mu$.

Since the applied field along $z$ and associated potential
is constant across the sample volume, this does not lead to any additional
scattering, which in the continuum limit happens only at discontinuities in the 
potential; still it must be included since it affects (or rather, effects) 
the relative phase of 
spin-flipped vs. non-spin-flipped portions of the neutron wavefunction, which 
changes the measured reflectivity.


\subsection{Reparametrization of $\psi$ and Reflectivity Derivation}
In the more general equation \ref{eq:mu_definition},
the values of $\mu_1$ or $\mu_3$ 
become unbounded when $\vec B$ approaches a direction perfectly parallel 
or antiparallel to the spin quantization direction $\hat z'$.
This situation of course always occurs in the fronting (and backing) medium since there the
field direction defines the quantization direction, $\hat z' \equiv \hat B_F$.
While the equations are analytically correct when one takes the appropriate limits,
floating-point computation errors are introduced when multiplying and dividing 
arbitrarily large numbers in a computer.

Since the $\mu$ values in Eq. \ref{eq:psi_z_definition} serve only to describe the 
ratio between the components of $\psi^+$ and $\psi^-$, 
and because $\mu_1 = \mu_2$ and $\mu_3 = \mu_4$,
we can rearrange that equation as
\begin{equation}
  \label{eq:reparametrized_psi}
  \begin{array}{l}
	\psi^{i,+}(z) = 
		D^i_1 e^{S^i_1 z} + 
		D^i_2 e^{S^i_2 z} +
		\gamma D^i_3 e^{S^i_3 z} +
		\gamma D^i_4 e^{S^i_4 z} \\
    \psi^{i,-}(z) = 
		\beta D^i_1 e^{S^i_1 z} + 
		\beta D^i_2 e^{S^i_2 z} +
		D^i_3 e^{S^i_3 z} +
		D^i_4 e^{S^i_4 z} \\
  \end{array}
\end{equation}
and relating these constants to our previous parametrization we get
\begin{equation}
  \begin{array}{l}
  	\beta = \mu_1 \\
  	\gamma = 1/\mu_3 \\
  	D_1 = C_1 \\
  	D_2 = C_2 \\
  	D_3 = C_3/\gamma \\
  	D_4 = C_4/\gamma
  \end{array}
\end{equation}

This solution to the S.E. is valid within any layer of the material, and
so we can calculate the reflectivity by using the boundary conditions to 
stitch together solutions from adjacent layers.  At any interface, the value of
the wavefunction and its first derivative $[\psi, \psi']$ must be continuous across
that boundary.  We can write the 
wavefunction in terms of the $D_j^i$ coefficients in that layer (for either incident spin state $i$):
\begin{equation}
\label{eq:cd_to_pz}
  \begin{pmatrix}
    \psi^{i,+}(z) \\
    \psi^{i,-}(z) \\
    \psi'^{i,+}(z) \\
    \psi'^{i,-}(z) \\
  \end{pmatrix}
  = \chi(z)
  \begin{pmatrix}
    D_1^i \\
    D_2^i \\
    D_3^i \\
    D_4^i \\
  \end{pmatrix}
\end{equation}
where from  Eq. \ref{eq:reparametrized_psi}:
\begin{equation}
  \label{eq:chi}
  \begin{array}{ll}
  \chi(z) &= 
  \begin{pmatrix}
    1 & 1 & \gamma & \gamma \\[0.3em]
    \beta  & \beta &  1 &  1 \\[0.3em]
    S_1  & -S_1  & \gamma & -\gamma \\[0.3em]
    \beta S_1  & -\beta S_1  & S_3  & -S_3  \\[0.3em]
  \end{pmatrix}
  \begin{pmatrix}
    e^{S_1 z} & 0 & 0 & 0 \\[0.3em]
    0 & e^{-S_1 z} & 0 & 0 \\[0.3em]
    0 & 0 & e^{S_3 z} & 0 \\[0.3em]
    0 & 0 & 0 & e^{-S_3 z} \\[0.3em]
  \end{pmatrix}
  \\[1.5em]
  {} &=
  \begin{pmatrix}
    1 & 1 & \gamma & \gamma \\[0.3em]
    \beta  & \beta &  1 &  1 \\[0.3em]
    S_1  & -S_1  & \gamma & -\gamma \\[0.3em]
    \beta S_1  & -\beta S_1  & S_3  & -S_3  \\[0.3em]
  \end{pmatrix}
  (e^{\mathbf{S}z} \cdot \mathbf{I})
  \end{array}
\end{equation}
where $\gamma$, $\beta$ and $\mathbf{S}$ are specific to the layer $l$ and incident spin state $i$
being calculated.
At the boundary between layers $l,l+1$ (we will define the boundary position $z\equiv Z_l$ here)
we have $\psi_l = \psi_{l+1}$ and $\psi'_l = \psi'_{l+1}$, so that
\begin{equation}
\chi_l(Z_l)
  \begin{pmatrix}
    D_{1,l} \\
    D_{2,l} \\
    D_{3,l} \\
    D_{4,l} \\
  \end{pmatrix}
= \chi_{l+1}(Z_l)
  \begin{pmatrix}
    D_{1,l+1} \\
    D_{2,l+1} \\
    D_{3,l+1} \\
    D_{4,l+1} \\
  \end{pmatrix}
\end{equation}
so to get $\{D_{l+1}\}$ from $\{D_l\}$, we invert $\chi_{l+1}$ and
\begin{equation}
\chi^{-1}_{l+1}(Z_l) \chi_l(Z_l)
  \begin{pmatrix}
    D_{1,l} \\
    D_{2,l} \\
    D_{3,l} \\
    D_{4,l} \\
  \end{pmatrix}
= 
  \begin{pmatrix}
    D_{1,l+1} \\
    D_{2,l+1} \\
    D_{3,l+1} \\
    D_{4,l+1} \\
  \end{pmatrix}
\end{equation}
where the formula for $\chi^{-1}$ can be calculated to be \footnote{
$\gamma$ and $\beta$ never have the same complex phase, so the 
denominator of Eq. \ref{eq:chi_inv} is never zero.
} 
\begin{equation}
\label{eq:chi_inv}
\chi^{-1}(z) = 
\frac{1}{2(1- \gamma\beta)}
  (e^{-\mathbf{S} z}\cdot \mathbf{I})
  \begin{pmatrix}
     1 & -\gamma & \frac{1}{S_1} & \frac{-\gamma}{S_1}   \\[0.3em]
     1 & -\gamma & \frac{-1}{S_1} & \frac{\gamma}{S_1} \\[0.3em]
    -\beta & 1 & \frac{-\beta}{S_3} & \frac{1}{S_3} \\[0.3em]
    -\beta & 1 & \frac{\beta}{S_3} & \frac{-1}{S_3} \\[0.3em]
  \end{pmatrix}
\end{equation}
Then for a structure with $N$ layers, the coefficients of the transmitted wave $\{D_{j,N}^i\}$ 
are related to the coefficients in the incident medium $\{D_{j,0}^i\}$ by 
\begin{equation}
  \label{eq:b_matrix_def}
  \begin{pmatrix}
    D_{1,N}^i \\
    D_{2,N}^i \\
    D_{3,N}^i \\
    D_{4,N}^i \\
  \end{pmatrix} = 
  \prod_N^{1} (\chi^i_n)^{-1}(Z_{n-1}) \chi^i_{n-1}(Z_{n-1})
  \begin{pmatrix}
    D_{1,0}^i \\
    D_{2,0}^i \\
    D_{3,0}^i \\
    D_{4,0}^i \\
  \end{pmatrix} = B^i
  \begin{pmatrix}
    D_{1,0}^i \\
    D_{2,0}^i \\
    D_{3,0}^i \\
    D_{4,0}^i \\
  \end{pmatrix}
\end{equation}
where the pairs of $\chi^{-1}_{n}(Z_{n-1}) \chi_{n-1}(Z_{n-1})$ are $4\times 4$
matrices.  Note that the matrices differ for the different incident spin states,
and so we have to calculate the matrix product $B^+$ and $B^-$ separately.
The remaining boundary conditions are met by identifying 
the coefficients in the fronting medium for the two polarized incident states
$I^+,I^-$
\begin{equation}
  \begin{pmatrix}
    D_{1,0} \\
    D_{2,0} \\
    D_{3,0} \\
    D_{4,0} \\
  \end{pmatrix}^+\!\!\! = 
  \begin{pmatrix}
    I^+ \\
    r^{+,+} \\
    0 \\
    r^{+,-} \\
  \end{pmatrix}\,\mathrm{and}\,
  \begin{pmatrix}
    D_{1,0} \\
    D_{2,0} \\
    D_{3,0} \\
    D_{4,0} \\
  \end{pmatrix}^{-}\!\!\! = 
  \begin{pmatrix}
    0 \\
    r^{-,+} \\
    I^- \\
    r^{-,-} \\
  \end{pmatrix}
\end{equation}
and the coefficients in the backing medium:
\begin{equation}
  \begin{pmatrix}
    D_{1,N} \\
    D_{2,N} \\
    D_{3,N} \\
    D_{4,N} \\
  \end{pmatrix}^+\!\!\! = 
  \begin{pmatrix}
    t^{+,+} \\
    0 \\
    t^{+,-} \\
    0 \\
  \end{pmatrix}\,\mathrm{and}\,
  \begin{pmatrix}
    D_{1,N} \\
    D_{2,N} \\
    D_{3,N} \\
    D_{4,N} \\
  \end{pmatrix}^-\!\!\! = 
  \begin{pmatrix}
    t^{-,+} \\
    0 \\
    t^{-,-} \\
    0 \\
  \end{pmatrix}
\end{equation}
Note that $D_{2,N}, D_{4,N}$ are zero because of the boundary condition that
the upward-traveling wave coefficient in the backing medium is zero (only downward-traveling
waves corresponding to transmission are physical in our experimental
setup.)

For the $I^+$ incident state, $I^-\equiv 0$ and vice versa, and so we can calculate the ratios
$r^{+,+} \equiv \frac{r^+}{I^+}$, $r^{+,-} \equiv \frac{r^-}{I^+}$, etc. from the $B$ matrix product of 
Eq. \ref{eq:b_matrix_def} by using the zeros in $D_{2,N}, D_{4,N}$, which gives two 
equations with two unknowns $(r^+, r^-)$ if we take the incident intensity to be unity.
This gives for the different cross sections
\begin{equation}
  \label{eq:r_from_b}
  \begin{array}{l}
    r^{+,+} = \frac{B^+_{24}B^+_{41} - B^+_{21}B^+_{44}}{B^+_{44}B^+_{22} - B^+_{24}B^+_{42}}\\[0.6em]
    r^{+,-} = \frac{B^+_{21}B^+_{42} - B^+_{41}B^+_{22}}{B^+_{44}B^+_{22} - B^+_{24}B^+_{42}}\\[0.6em]
    r^{-,+} = \frac{B^-_{24}B^-_{43} - B^-_{23}B^-_{44}}{B^-_{44}B^-_{22} - B^-_{24}B^-_{42}}\\[0.6em]
    r^{-,-} = \frac{B^-_{23}B^-_{42} - B^-_{43}B^-_{22}}{B^-_{44}B^-_{22} - B^-_{24}B^-_{42}}\\[0.6em]
  \end{array}
\end{equation}

As can been seen in Eq. \ref{eq:chi} above, 
the new constants $\gamma$ and $\beta$ have real physical significance as the
mixing terms between $\psi^+$ and $\psi^-$ in a given layer, and for 
any $B_{z'} \geq 0$ the constants $\gamma$ and $\beta$ are found inside the unit 
circle in the complex plane, i.e. $|\gamma, \beta| \leq 1$.
In the fronting and backing media, they are both identically
zero.

For a layer perfectly antiparallel to $\hat z'$, $\beta$ and $\gamma$ will still be unbounded, but we
further note that the numbering of the roots in Eq. \ref{eq:propagation_constants}
is arbitrary, so for every layer where $B_{z'} < 0$ we perform this switch for the matrix
corresponding to that layer:
$S_1' \rightarrow S_3$,
$S_3' \rightarrow S_1$, $\gamma' \rightarrow 1/\beta$
and $\beta' \rightarrow 1/\gamma$.  The new $\beta'$ and $\gamma'$ again have a magnitude 
less than or equal to one, 
and we can carry on with the calculation.  
This has no effect on the calculated reflectivity
\footnote{
However if the calculated values of $D_j$ are to be used to reconstruct
the full wavefunction within that layer (say, for a Distorted-Wave Born Approximation calculation)
one has to be aware of the switch that was made, so that the multiplier $D_j$ is 
correctly associated with the propagation vector $S'_j$ instead of $S_j$.
},
and the matrices are now 
all well-conditioned (the magnitude of matrix elements is always 
less than or equal to one.)  As in the parallel case, for perfectly antiparallel $\vec B$
the mixing terms are exactly zero. 


It is interesting that in this new parametrization, the degenerate case where the magnetization 
is always aligned parallel or antiparallel to the applied $\vec H$ reduces very obviously
to two uncoupled equations for the propagation of $\psi^+$ and $\psi^-$, since the
mixing terms in every layer are zero.  

Since the spin of the incoming beam is never
flipped in this case, the reference energy (including a Zeeman term)
for the reflected neutron in the fronting medium will match the energy of the incident neutron for both possible 
incident spin states, and it can be subtracted from all the equations with no effect as an arbitrary energy offset.  
Thus the Zeeman correction to the expected reflectivity will only be needed when there is non-collinear
magnetization of the layers, but when this correction has to be made it will alter
all the cross-sections, including the non-spin-flip reflectivity (because of cross-terms in the calculation
between spin-flip and non-spin-flip reflectivity).

\subsection{Parametrization of $k$, $E$}
The wave propagation constants $S$ in Eq. \ref{eq:propagation_constants} are dependent
only on the fixed potentials $\rho_B, \rho_N$ for that layer, and an energy term which depends
on the spin state and $k_z$ of the incident neutron.  If the reflectivity is solved
for a given $E$, this corresponds to a set of $k_{F,z}^{+} \neq k_{F,z}^{-}$:
\begin{equation}
	k_{F,z}^{\pm} = \sqrt{\frac{2m}{\hbar^2}E - 4\pi(\rho_{F,N} \pm \rho_{F,B}) }
\end{equation}

While this saves roughly a factor of two in computation time by mapping a single
energy to the corresponding $k$ for the two incident spin states, it does not match
the way a reflectometry experiment is typically carried out, where all 4 spin-dependent
cross-sections are measured for a single incident wavevector.  A more
natural instrument coordinate system is based on the incident and reflected angles
$(\theta_\textrm{in}, \theta_\textrm{out})$, which maps onto $(k_\textrm{in}, k_\textrm{out})$,
and so we calculate the reflectivity twice for each value of $k_\textrm{in}$, once for
each spin state and corresponding value of $E_{F,z}^i$. 

%

%
%
%
%

\section{Measurement setup}
\subsection{Sample and detector angles}
While the shift in the reference potential had a large effect on the calculated
reflectivities above, it is the angular shift
(i.e., $\theta_\textrm{out} - \theta_\textrm{in}$)
in the spin-flipped reflected beams
that most affects the instrument setup for this type of measurement.

From the shift in $k_z$ in Eqs. \ref{eq:pm_shift} and \ref{eq:mp_shift}, we can
calculate the outgoing angle of the reflected beam by
\begin{equation}
	\label{eq:outangle}
	\theta_\mathrm{out} = \arctan (k_{z(\mathrm{out})} / k_{x})
\end{equation}

From Eq.~\ref{eq:outangle}, it is easy to see that the angular shift of the spin-flipped 
reflected beams changes during the measurement, thus a position-sensitive neutron detector 
will clearly facilitate experiments when the Zeeman effect is significant. However, 
some existing reactor-based PNR beamlines use pencil detectors. Pencil detectors 
have their own advantage of very high detection efficiency, but an unconventional 
experimental procedure is required to take care of the Zeeman effect. 
Below we detail the experimental setup using a pencil detector when the Zeeman 
effect is significant.
For the four possible spin cross-sections, three different values of $k_{z(\mathrm{out})}$
(and therefore detector angle) are found for a single $k_{z(\mathrm{in})}$ 
in the specular condition $(k_{x(\mathrm{in})} = k_{x(\mathrm{out})})$; one spin-flipped state is shifted higher and the
other spin-flipped state is shifted lower, while the two non-spin-flip processes
give $k_{z(\mathrm{in})} = k_{z(\mathrm{out})}$, so that $\theta_\mathrm{in} = \theta_\mathrm{out}$.
One could just as well choose a fixed $\theta_\mathrm{out}$ and $k_{z(\mathrm{out})}$,
and calculate the three possible values of $k_{z(\mathrm{in})}$ for specular scattering,
but for this discussion we will use $k_{z(\mathrm{in})}$ as the fixed quantity.

Since the polarization efficiency of the measurement system typically 
depends on the instrument geometry, for each of the three $\theta_\mathrm{out}$ 
corresponding to a specularly-reflected beam, all four
spin cross-sections have to be measured in order extract an efficiency-corrected reflectivity for that
angle.  Only one of the corrected reflectivities out of four will be used from the measurements
at Zeeman-shifted angles $\theta_\mathrm{out}^{-+}$ and $\theta_\mathrm{out}^{+-}$, 
while two reflectivities can be extracted from the non-spin-flip 
$\theta_\mathrm{out}^{++} = \theta_\mathrm{out}^{--} = \theta_\mathrm{in}$.
Overall this increases the measurement time by a factor of three compared to an experiment without 
Zeeman corrections.


\section{Example measurement}
\subsection{In-plane magnetic sample}
\begin{figure}
	\includegraphics[width=1.0\linewidth]{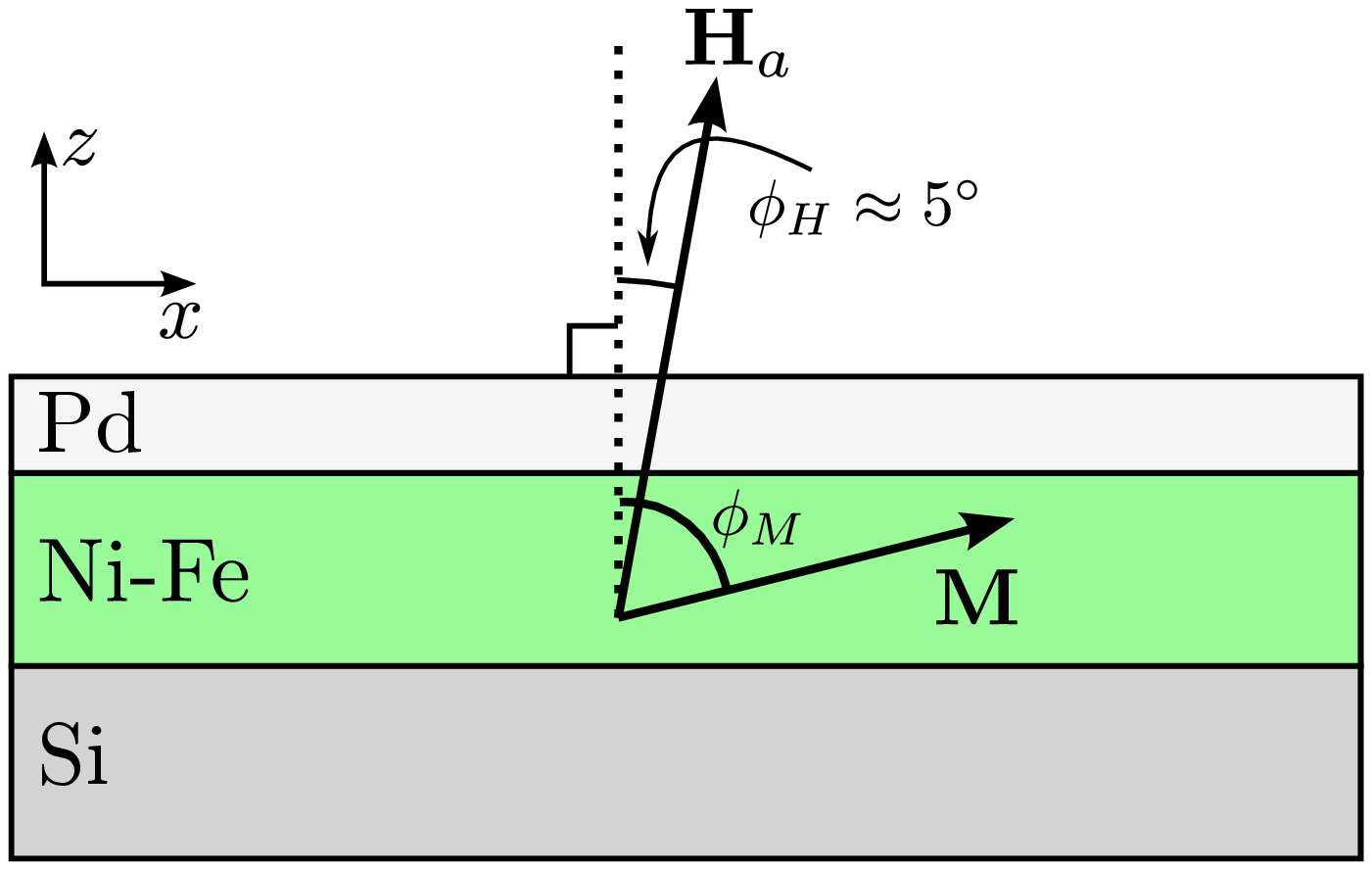}
	\captionof{figure}{Test sample: Side view of the layer structure of 
	Pd (200 \AA) on Ni-Fe alloy (600 \AA) on Si substrate.  
	Sample lateral size is 25 mm $\times$ 25 mm.
	The external applied field is slightly tilted with respect to the surface normal.}
	\label{fig:sample}
\end{figure}

In order to realize a large moment non-collinear with the field, a sample of a very magnetically
soft material (Ni-Fe alloy) was grown on a single crystal Si substrate, and capped
with a layer of Pd to prevent oxidation (as seen in Fig. \ref{fig:sample}).

For the principal polarized neutron reflectometry measurement of this study, 
the external magnetic field was applied for the measurement at a small angle to the film surface
normal as seen in the figure.  
The demagnetizing field (shape anisotropy) of the film acts to keep the magnetization in-plane,
and for appropriate choices of field strength and angle this dominates over the 
torque from the applied field, so that the
magnetization remains largely in-plane.  At the same time, the small in-plane component of the field 
$H_x$ is enough to align the layer into a single domain, pointing mostly along $x$.

This arrangement provides an ideal test of the equations, since there is both a large moment 
$\mathbf M \perp \mathbf H$ providing spin-flip scattering, 
and simultaneously a large $\mathbf H$ field which causes Zeeman splitting of those spin-flipped neutrons.

We verified with a vibrating-sample magnetometer measurement that the test sample is indeed
magnetically soft with a saturation field in the hard (out-of-plane) direction of about 0.5 T, 
and at 0.244 T (the applied field for the neutron measurements) the out-of-plane loop is linear
with field, suggesting a coherent rotation.  This verifies that it is a magnetically
soft film with the expected shape anisotropy and no significant domain formation at the neutron
measurement condition.

A supplementary reflectometry measurement of the same sample was done 
in an in-plane saturating field in order to get a good value of the saturation magnetization of the
soft magnetic layer.  The scattering results from this measurement (not shown) 
are easily fit to standard models
of polarized neutron reflectometry without Zeeman corrections and indicate a saturation internal 
$B$-field of 0.551 T ($M = 439.53$ kA/m) \footnote{
This is below the expected value for a Ni-Fe alloy, which may result from the incorporation of oxygen
in the film due to a poor
vacuum during the deposition process.  For the purposes of this investigation all that is required is a magnetically
soft film and the exact magnetization is irrelevant.}.

\subsection{Results}
The reflectivity measurements were undertaken at the Polarized Beam Reflectometer instrument (PBR)
at the NIST Center for Neutron Research, with supermirror spin polarizer and analyzer and
current-coil Mezei-type spin flippers for the incident and reflected beams.  
In an applied field $\mu_0 H_a=244$ mT at an angle as shown in Fig. \ref{fig:sample},
for a series of 
$k_{z(\mathrm{in})}$, all four spin cross-sections were 
measured at each of the the three outgoing angles corresponding to 
$(k_{z,F}^{+-})$, $(k_{z,F}^{+-})$, and $(k_{z,F}^{++},k_{z,F}^{--})$.  The data for each of those
outgoing angles was polarization-corrected and the relevant cross-sections were extracted. 

\begin{figure}
	\includegraphics[width=1.0\linewidth]{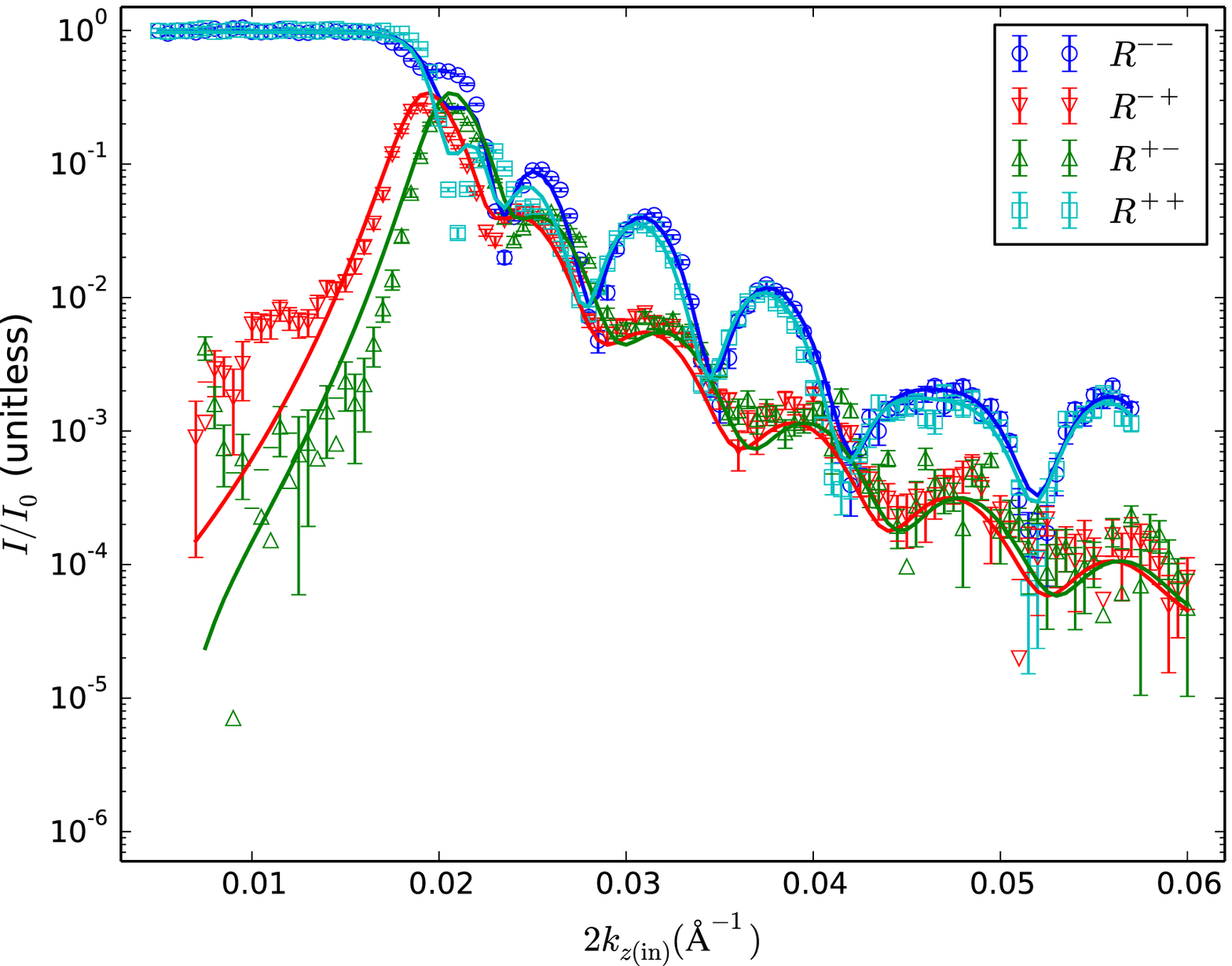}
	\captionof{figure}{Reflectivity of test sample without including the effects of the Zeeman energy.
	Data is open symbols, with error bars corresponding to $\pm 1\sigma$ 
	according to counting statistics and resolution function of the instrument; fits are
	the solid lines (reduced $\chi^2$ for this fit is 25.0).
	Data is parametrized and fit according to $k_{z(\mathrm{in})}$.}
	\label{fig:R4_noZeeman}
\end{figure}

First in Fig. \ref{fig:R4_noZeeman} we show a the best fit to the data performed using the freely-available 
Refl1D \cite{KirbyCOCIS, refl1d} package, but without making corrections for the Zeeman effects.
The symbols represent data points with error bars and the 
lines represent the best fit possible.

\begin{figure}
	\includegraphics[width=1.0\linewidth]{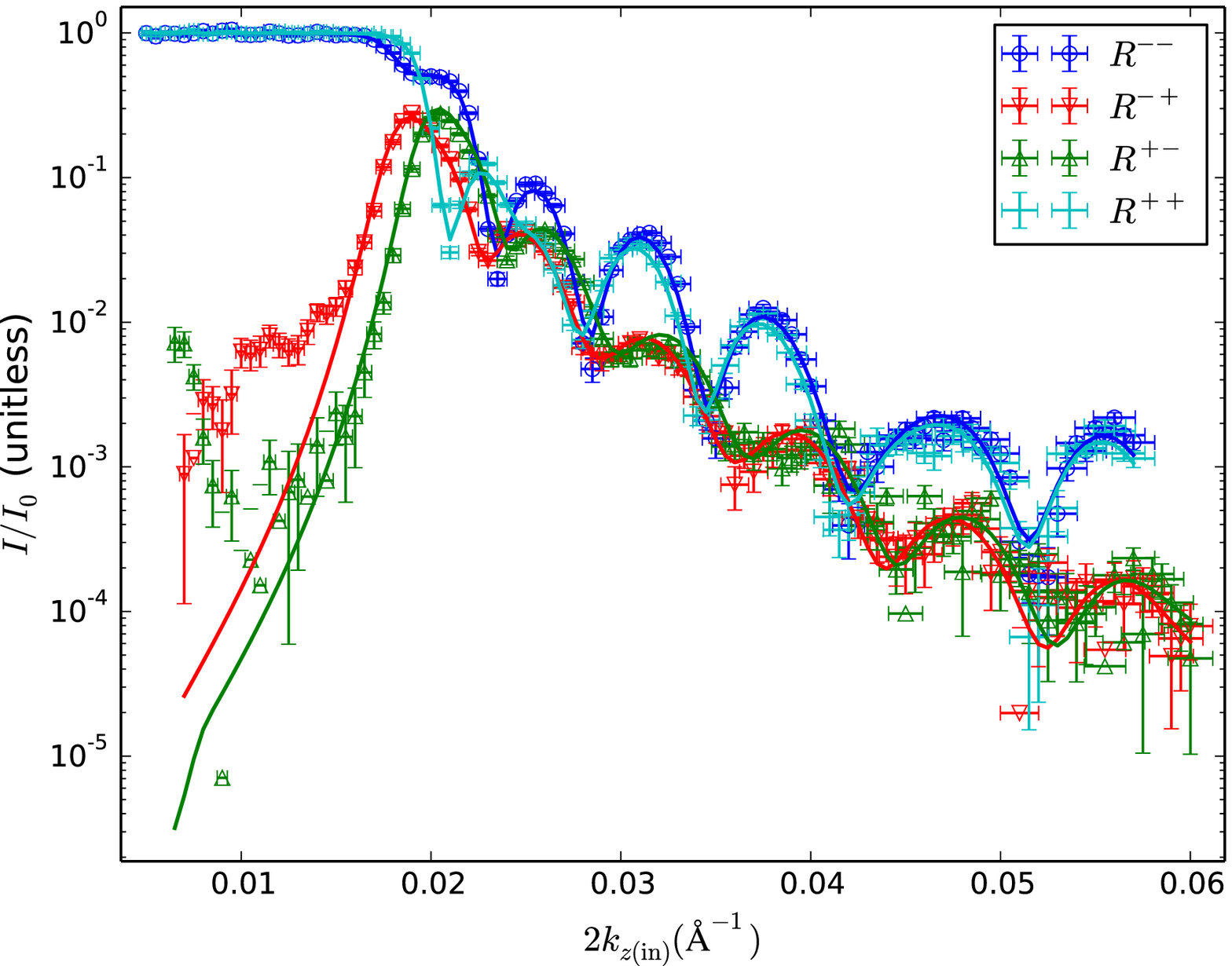}
	\captionof{figure}{Reflectivity of test sample, in all 4 cross sections, including fit.
	Data is open symbols, with error bars corresponding to $\pm 1\sigma$ 
	according to counting statistics and resolution function of the instrument; fits are
	the solid lines (reduced $\chi^2$ for this fit is 3.7).
	Data is parametrized and fit according to $k_{z(\mathrm{in})}$.}
	\label{fig:R_4xs}
\end{figure}

We compare this to a fit performed using a modification of the software which includes
the changes to the theory described in the first section of this
manuscript.  Both the data and the fit are presented in Fig. \ref{fig:R_4xs}.

\begin{figure}
	\includegraphics[width=1.0\linewidth]{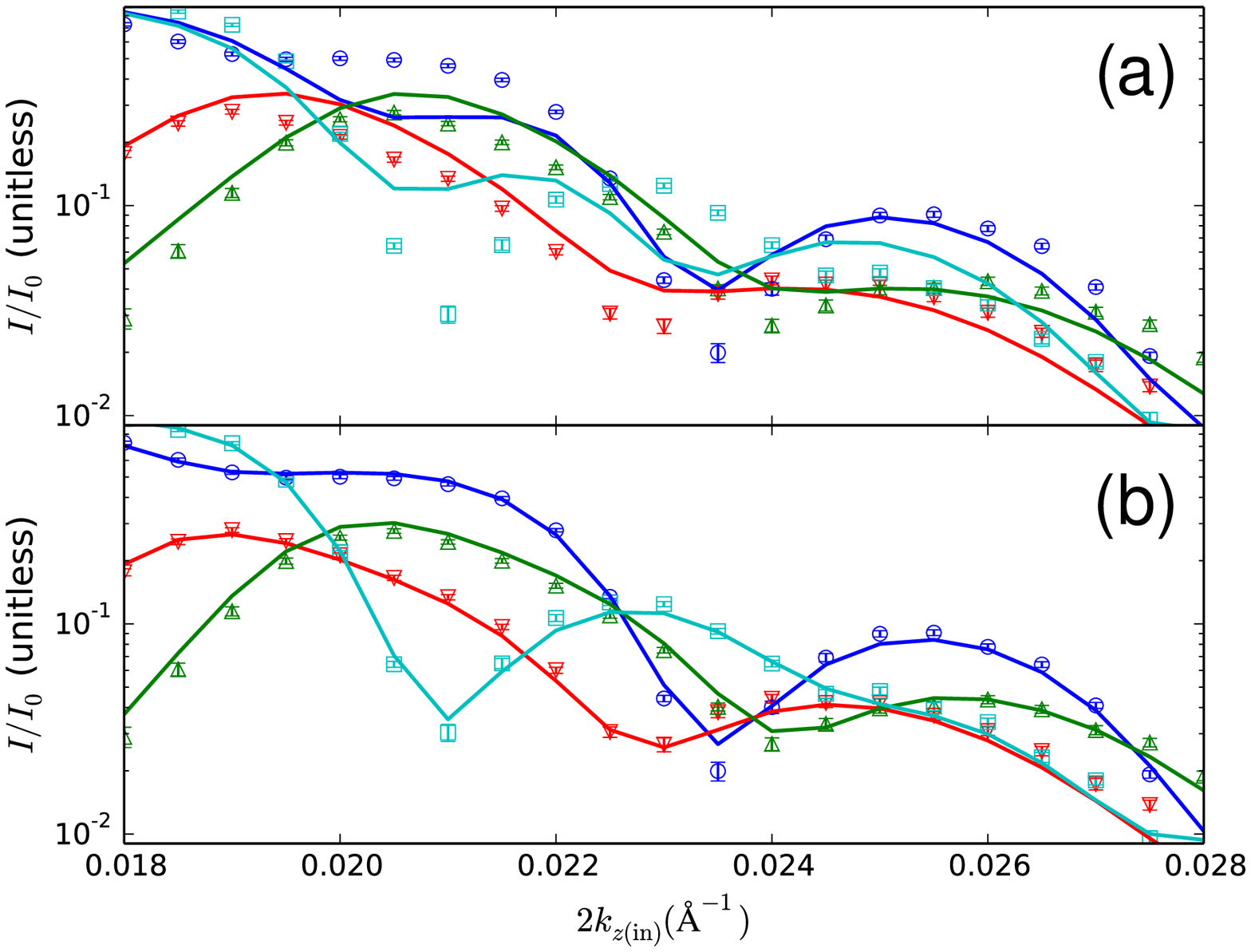}
	\captionof{figure}{Enlargement of the reflectivity fits near the critical edge.
	(a) corresponds to the fit without Zeeman corrections in Fig. \ref{fig:R4_noZeeman}
	and (b) corresponds to the corrected fit in Fig. \ref{fig:R_4xs}.
	A clear improvement in the quality of the fit is seen.
	Symbols and lines have the same meanings in this plot as in the originals.}
	\label{fig:R4_blowup}
\end{figure}

In the uncorrected fit in Fig. \ref{fig:R4_noZeeman} 
we can clearly see that the splitting between the non-spin-flip scattering 
at low $k_{z\mathrm{(in)}}$ is grossly underestimated in the best-fitting model.  In this region the error bars are small
due to the strong scattering and this is what leads to the large minimum $\chi^2$ value of 25.0 for this 
fit.  An enlargement of this region for comparing corrected vs. uncorrected fits is show in Fig. \ref{fig:R4_blowup}.

By contrast the Zeeman-corrected fit is very good, with a chi-squared value of 3.7.  
The visible deviations of the spin-flip data from the fit
at very low $k_{z\mathrm{(in)}}$ are likely due to issues with the polarization correction (the correction
is of the same magnitude as the spin-flip data there), and this does not significantly affect the rest of the fit.
In the enlarged plot in Fig. \ref{fig:R4_blowup}(b) this fit clearly reproduces the data near the critical edge.
The best fit to the data corresponds to a magnetic scattering length density in the Ni-Fe layer of 
$\rho_B=1.12\times 10^{-6} \textrm{\AA}^{-2}$ and thus 
$M_x = 385$ kA/m.  

\begin{figure}
	\includegraphics[width=1.0\linewidth]{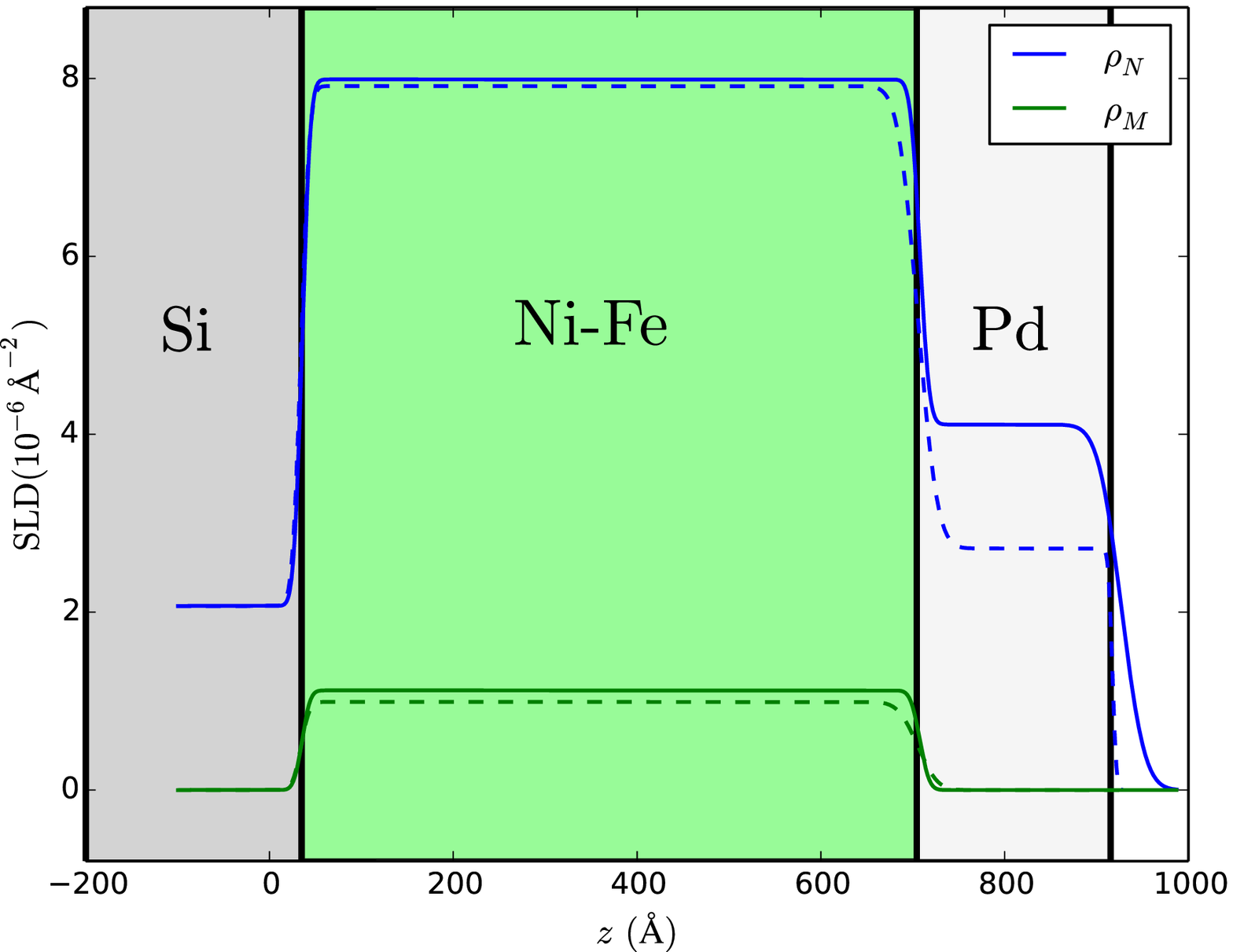}
	\captionof{figure}{Scattering length density profile (SLD)
	corresponding to the fits shown in Fig. \ref{fig:R_4xs} (solid line) and 
	Fig. \ref{fig:R4_noZeeman} (dashed line).  $\rho_N$ and $\rho_M$ refer to the nuclear
	and magnetic scattering length densities respectively in blue and green. Profile is overlaid
	on a color-coded diagram of the sample structure as seen in Fig. \ref{fig:sample} 
	for reference.}
	\label{fig:profile}
\end{figure}

The SLD profiles resulting from the fits in Figs. \ref{fig:R4_noZeeman} and \ref{fig:R_4xs}
(nuclear and magnetic) are shown in Fig. \ref{fig:profile} with dotted lines, 
while the SLD profiles from the corrected fit are shown with solid lines.  
The difference between the profiles is most prevalent in the region of
the capping layer, where the uncorrected fit gives an unphysically low value of nuclear 
SLD of the Pd capping layer (2.7 \AA$^{-1}$ rather than the expected value of 4.1 \AA$^{-1}$,)
and an unrealistically low roughness for the top interface, where one would expect the top interface to have similar roughness
to the interface immediately below.

The out-of-plane component of $\vec M$ for a system with uniaxial anisotropy arising from the demagnetization
field is expected to be linearly dependent (when coherently rotating across the entire sample)
on an applied out-of-plane field, reaching the saturation 
value at $H_k = 4\pi M_S$.  In our case $M_z \approx (H_a / H_k)M_S = (0.244/0.551)M_S$, and
since $M_z = M_S\cos\phi$ and $M_x = M_S\sin\phi$, we can extract an expected value of the 
in-plane magnetization $M_x \approx 394$ kA/m, which agrees well with the
fit value of 385 kA/m.

The most striking feature of the scattering in Fig. \ref{fig:R_4xs} is the large splitting between
the non-spin-flip reflectivities $R^{++}, R^{--}$ at low $k_{z(\mathrm{in})}$, but which 
disappears at higher $k_z$.  This is a signature of the Zeeman effect, which will be most
pronounced when the Zeeman energy is comparable to the kinetic energy along the scattering direction.

The best indication that this splitting is a result of the Zeeman effect is to compare with data fitted
to a model with no Zeeman energy included; this is shown in Fig. \ref{fig:R4_noZeeman}.

In Figs. \ref{fig:R_4xs} and \ref{fig:R4_noZeeman}
there is an apparent horizontal shift between the two spin-flip reflectivities.  
This is entirely due to the choice of
plotting that data as a function of $2k_{z(\mathrm{in})}$.  If we had chosen to plot vs. the total momentum
transfer $Q$ the features would be mostly aligned, but the advantage of plotting it this
way is that the scattering sum rules are more apparent; for an incident beam $I^-$ at low angles
where the scattering is strong, we can clearly see the non-spin-flip reflectivity $R^{--}$ 
has a dip when $R^{-+}$ has a peak (a similar correspondence is seen between $R^{++}$ and $R^{+-}$.)

\section{Conclusions}

We have described a procedure for measuring polarized neutron reflectivity in high fields, including
important changes to the modeling and instrument configuration due to Zeeman shifts in the
energy and angle of spin-flip scattered neutrons.  These considerations will be important for 
characterization of thin films with large magnetic anisotropy, which are a component of a growing
number of technologically relevant systems.

A data-modeling package with the necessary modifications for this type of measurement was demonstrated
to provide accurate quantitative fits of a test system, and this software is now readily available to
the research community \cite{refl1d}.  The deviations
from non-Zeeman-corrected polarized specular neutron modeling are most pronounced where the spin-flip 
scattering is most intense.




\ack{Acknowledgements}
Y. Liu is supported by the Division of Scientific User Facilities of the Office of Basic Energy Sciences, US Department of Energy.

\bibliographystyle{iucr} 
\bibliography{extracted}



\end{document}